\documentclass[runningheads]{llncs}
\usepackage{subfig}
\usepackage{graphicx}
\usepackage{float}

\usepackage[table,xcdraw]{xcolor}
\begin{document}
\title{Is knowledge the key? An experiment on debiasing architectural decision-making - a pilot study}
\titlerunning{Debiasing architectural decision-making -- a pilot study}
%
\author{Klara Borowa \inst{1}\orcidID{0000-0002-7160-5950} \and
Robert Dwornik \inst{1}\and
Andrzej Zalewski\inst{1}\orcidID{0000-0001-5254-4761}}
%
%
\institute{Warsaw University of Technology, Institute of Control and Computation Engineering, Warsaw, Poland \\
\email{klara.borowa@pw.edu.pl}}
%
\maketitle              
\begin{abstract}
The impact of cognitive biases on architectural decision-making has been proven by previous research. In this work, we endeavour to create a debiasing treatment that would minimise the impact of cognitive biases on architectural decision-making. We conducted a pilot study on two groups of students, to investigate whether a simple debiasing presentation reporting on the influences of cognitive biases, can provide a debiasing effect. The preliminary results show that this kind of treatment is ineffective. Through analysing our results, we propose a set of modifications that could result in a better effect. 


\end{abstract}
\section{Introduction}
The occurrence of cognitive biases is inherent to the human mind, and as such, can influence all individuals taking part in the software development process \cite{Mohanani2018}: developers
\cite{Chattopadhyay2020}, architects \cite{Tang2011}, designers \cite{Mohanani2014}, testers \cite{Calikli2013a}.

In particular, cognitive biases have been proven to distort architectural decision-making \cite{VanVliet2016} by influencing software architects' reasoning \cite{Tang2011}. This influence can be particularly strong, since every systems architecture is actually a set of design decisions \cite{Jansen2005} made by individuals. Thorough education about cognitive biases turned out to significantly improve software effort estimation \cite{Shepperd2018}, which is severely afflicted by cognitive biases \cite{halkjelsvik2018time}. Similarly, in this work we examine, \textit{(RQ) whether educating software architects about cognitive biases can provide a beneficial debiasing effect, which increases the rationality of decision-making}. 

In order to answer this question, we designed an experiment and ran a pilot study on two groups of students. The preliminary findings show that educating engineers about the possible impact of cognitive biases is not sufficient to  mitigate the influence of cognitive biases on design decisions. 

Therefore, more advanced debiasing techniques are needed.
We analysed how exactly cognitive biases influenced various elements of the conversation (arguments, counterarguments, and general conversation). Based on that, we proposed additional debiasing techniques that can be used in order to create a more effective debiasing treatment. We plan to perform a modified version of this experiment, on a larger sample, in the near future. Our long time objective is to develop effective, debiasing techniques for architectural decision-making.

\section{Related Work} 
The concept of cognitive biases was introduced by Tversky and Kahneman in their work about Representativeness, Availability and Anchoring biases \cite{tversky1974judgment}. Cognitive biases are a by-product of the dual nature of the human mind -- intuitive (known as System 1) and rational (known as System 2) \cite{kahneman2011thinking}. When the logic-based reasoning of System 2 is not applied to the initial decisions of System 1, we can say that the decision was biased.

Software architecture, defined as set of design decisions \cite{Jansen2005}, is influenced by various human factors \cite{Tang2017}. One of these factors are cognitive biases \cite{VanVliet2016}. Their influence on architectural decision-making has been shown as significant in recent  research \cite{Tang2011} \cite{VanVliet2016} \cite{Manjunath2018} \cite{Zalewski2017}. When no debiasing interventions are applied, the consequences of such biased decisions can be severe -- for example resulting in taking on harmful Architectural Technical Debt \cite{Borowa2021}.

In the domain of architecture decision-making, various debiasing techniques were proposed \cite{VanVliet2016}, \cite{Borowa2021}. The use of techniques that prompt designers to reflect on their decisions, have turned out to be effective in improving the quality of the reasoning behind design decisions \cite{Tang2018}.

Debiasing, by educating software developers about the existence of cognitive biases and their influences, has recently been proven to work as a powerful tool in the realm of software effort estimation \cite{Shepperd2018}. The effectiveness of this approach to debiasing architectural decision making, has not yet been empirically tested.

\section{Study Design}
\label{study_design_section}
\subsection{Bias selection}
Based on the cognitive biases researched previously in relation to software development \cite{Mohanani2018}, as well as biases shown previously as influencing software architecture \cite{Zalewski2017}, \cite{Borowa2021}, \cite{VanVliet2016}, we selected three cognitive biases as the subject of the experiment:
\begin{enumerate}
    \item Anchoring -- when an individual over-relies on a particular solution, estimate, information or item, usually, the first one that they discovered or came up with \cite{tversky1974judgment}.
    \item Optimism bias -- when baseless, overly positive estimates, assumptions and attributions are made \cite{Ralph2011}.
    \item Confirmation bias -- the tendency to avoid the search for information that may contradict one's beliefs \cite{Stacy1995}.
\end{enumerate}

\subsection{Data acquisition}
In order to obtain the data for our study, we took part in four meetings with two groups of students that were working on a group project during their coursework. The meetings were conducted online through the MS Teams platform. Both groups were supposed to plan, design and implement a system as a part of their course. The topic for the project was at their discretion, with the only hard requirement being the use of Kubernetes in their solution.

In the case of one of the groups, we prepared a presentation during which we explained the concept of cognitive biases, and how they can influence architectural decision-making. We explicitly explained the three researched cognitive biases and gave examples of their possible influence on the students' project. We did not mention anything about cognitive biases or debiasing to the second group.

The meetings proceeded as follows:
\begin{enumerate}
    \item We asked the participants for their consent to record the meeting and to use their data for the purpose of our research.
    \item In the case of the debiased group (Team 2), we showed them our presentation about cognitive biases in architectural decision-making. We did not perform this action with the other group (Team 1).
    \item The meeting continued naturally, without our participation, although a researcher was present and made notes when necessary.
\end{enumerate}

We also asked the participants to fill in a small survey to obtain basic statistical data about them.

\subsection{Data Analysis}
The recordings from the meetings were transcribed. In order to identify the cognitive biases, and their influence on decision-making, we defined a coding scheme presented in Table \ref{tab:coding_table}. The codes were applied to indicate the occurrence of the researched biases, as well as the arguments for and against the discussed architectural decisions.

\begin{table}[]
\centering
\begin{tabular}{|l|l|l|}
\hline
Code   category &
  Code &
  Definition \\ \hline
Bias -- Anchoring &
  KOT &
  \begin{tabular}[c]{@{}l@{}}Putting too much emphasis on the first piece of \\      information or idea that was heard/proposed/invented.\end{tabular} \\ \hline
Bias -- Optimism &
  OPT &
  \begin{tabular}[c]{@{}l@{}}Naive   faith that the unpleasant consequences of our decisions \\      will not happen. Typical statements include: ``It will somehow be.''\\ ``No need to think about possible problems.'', ``Let's just start coding, \\ it will be fine.''\end{tabular} \\ \hline
Bias -- Confirmation &
  POT &
  \begin{tabular}[c]{@{}l@{}}Not   accepting and not seeking     information that is inconsistent with our  \\ current beliefs.\end{tabular} \\ \hline
Arguments for   the decision &
  ARG &
  \begin{tabular}[c]{@{}l@{}}An argument that was in favour of choosing a particular solution.\end{tabular} \\ \hline
Arguments   against the decision &
  PARG &
  \begin{tabular}[c]{@{}l@{}}A counterargument, against choosing a particular solution.\end{tabular} \\ \hline
\end{tabular}
\caption{Coding Scheme}
\label{tab:coding_table}
\end{table}

The first and second author coded the transcripts independently. Then, they used the negotiated coding \cite{Garrison2006} method to discuss and correct the coding until they reached a full consensus.

Subsequently, we counted the number of occurrences of each code, and analysed the fragments of the meetings that were found to have been influenced by cognitive biases.

\subsection{Participants}
We recorded four meetings with two different groups of students that were working on their Master’s degrees in Computer Science at Warsaw University of Technology. The students grouped themselves into teams depending on their own preferences and had to choose a team leader. The teams consisted of five members each. Most of the students (with a single exception) had prior professional experience in software development. More detailed information on the students is presented in Table \ref{tab:participant-data}.
\begin{table}[h]
\centering
\resizebox{\textwidth}{!}{%
\begin{tabular}{|l |l |p{3cm} |l |l|l|}
\hline
{\textbf{Age}} &
  {\textbf{Gender}} &
  {\textbf{Has professional experience?}} &
  {\textbf{Job position}} &
  {\textbf{Experience {[}years{]}}} & {\textbf{Team No}} \\ \hline
{23} & {M} & {Yes} & {Data Engineer}           & {1} & {1 (not debiased)}   \\ \hline
{24} & {M} & {Yes} & {Software Developer}        & {2.5} & {1 (not debiased)}  \\ \hline
{23} & {M} & {Yes} & {Software Developer - intern}               & {0.1} & {1 (not debiased)}\\ \hline
{24} & {M} & {Yes} & {Cloud/DevOps}            & {3} & {1 (not debiased)}    \\ \hline
{23} & {M} & {Yes} & {Systems Engineer} & {2} & {1 (not debiased)}    \\ \hline

{24} & {M} & {Yes} & {Java Developer}      & {1.5} & {2 (debiased)}  \\ \hline
{24} & {M} & {Yes} & {Full Stack Developer} & {2} & {2 (debiased)}   \\ \hline
{24} & {M} & {Yes} & {Java Developer}     & {2} & {2 (debiased)}   \\ \hline
{23} & {F} & {Yes} & {Sales Analyst}       & {1} & {2 (debiased)}   \\ \hline
{25} & {M} & {No}  & {No professional experience}                    & {0} & {2 (debiased)}   \\ \hline
\end{tabular}%
}
\caption{Participant data}
\label{tab:participant-data}
\end{table}

\section{Results}
\label{results_section}
\begin{figure}[h]
    \centering
    \includegraphics[width=12cm]{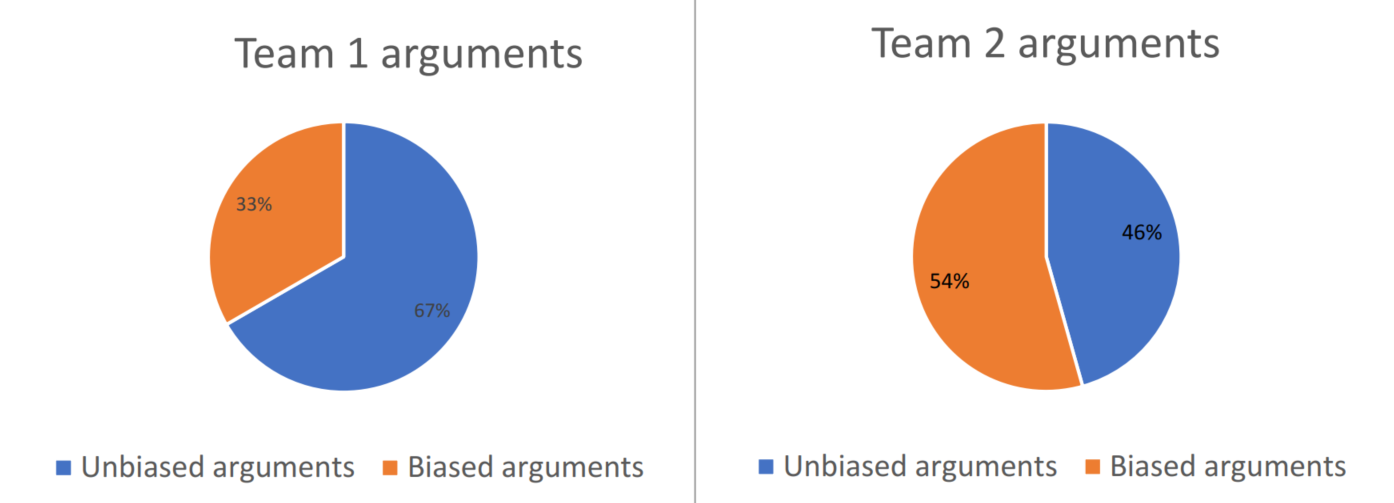}
    \caption{Biased arguments}%
    \label{fig:arguments_figure}%
\end{figure}

\begin{figure}[h]
    \centering
    \includegraphics[width=12cm]{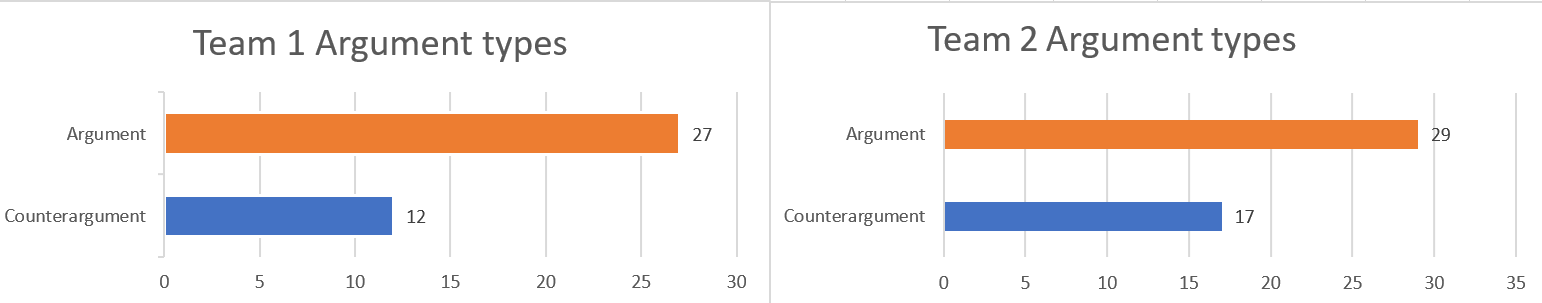}
    \caption{Argument count}%
    \label{fig:argument_count}%
\end{figure}

\begin{figure}[h]
    \centering
\includegraphics[width=15cm]{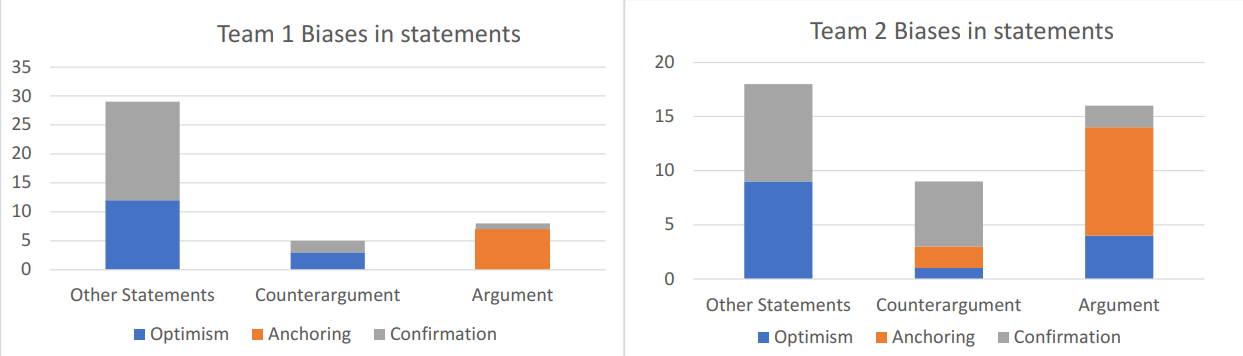} 
    \caption{Biases in statements}%
    \label{fig:biased_statements}%
\end{figure}

Using the coding scheme presented in Table \ref{tab:coding_table}, we obtained the following information:
\begin{itemize}
    \item The percentage of biased arguments in statements for or against certain architectural decisions (see Figure \ref{fig:arguments_figure}).
    \item How many arguments for and against certain architectural decisions were made during the meeting (see Figure \ref{fig:argument_count}).
    \item How many of these arguments and counterarguments, were influenced by cognitive biases (see Figure \ref{fig:biased_statements})
    \item How many cognitive biases were present in statements not related to architectural decisions (see Figure \ref{fig:biased_statements}).
\end{itemize}

Figure \ref{fig:arguments_figure}, which presents the percentage of biased arguments used during the meetings, shows that Team 1 (non-debiased) used more rational arguments than Team 2 (debiased). This means that the debiasing treatment – simply informing the participants about the existence of cognitive biases – was ineffective.  

Figure \ref{fig:argument_count}  shows that there was a significant difference between the amount of arguments and counterarguments in the discussions. Teams were less likely to discuss the drawbacks of their decisions than their positive aspects.

Figure \ref{fig:biased_statements} illustrates the number of biased statements, as well as the ratio between the researched biases depending on statement type.

In the case of both teams, most cognitive biases were present in statements not related to architectural decision-making. In this type of discussion, confirmation bias and optimism bias were the most prevalent. This was usually due to the teams’ need to reassure themselves that their course of action was correct.

In both teams, most of the biased arguments were influenced by the anchoring bias. This means that both teams considered an array of solutions that came to their minds first, without any additional argumentation on why the specific solution is correct. When it comes to counterarguments, against specific architectural solutions, confirmation bias was prevalent in both teams. This was usually due to the teams’ unwillingness to change a previously made decision.  

\section{Threats to validity}
In this work, we describe a pilot study. Its main weakness is the small number of participants that took part in the experiment. This means that all of our findings are preliminary and cannot be perceived as final. We plan to perform a modified version of this experiment with a larger number of teams, to obtain more data to verify our findings.

\section{Discussion}
The team that was not debiased by our presentation used a significantly lower number of biased arguments. This implies that a simple debiasing treatment, by simply reporting on the biases is not strong enough to counter the influence of cognitive biases on architectural decision-making.

We discovered the typical scenario of bias-influenced architectural decision making.
First, one team member proposes an idea that first came to their mind (an idea prompted by System 1). If the solution does not disturb the current project, other team members are unlikely to give any counterarguments (only around half of the arguments used were counterarguments) as they are already anchored on the initial proposition. If the solution requires changes to previously made decisions, other team members (due to confirmation bias), are likely to give biased counterarguments to avoid changes.
Additionally, the whole atmosphere of the conversation is heavily influenced by the confirmation bias and optimism bias, making the team unlikely to notice any errors in their decision-making.

With these findings in mind, we propose (Section \ref{research_outlook_section}) a set of modifications to our debiasing approach.

\section{Research outlook}
\label{research_outlook_section}
Since the pilot study showed that a simple debiasing treatment does not help to overcome the biases, we plan to extend and repeat this experiment with the following modifications:
\begin{itemize}
    \item Since the most biased arguments in favour of a solution were influenced by anchoring, and participants were overall less likely to use counterarguments -- we propose that the person presenting a solution, should also present at least one drawback.
    \item Since most biased counterarguments were influenced by confirmation bias, due to the teams’ reluctance to change a previously made decision -- we propose that one of the team members should monitor the discussion and point out the occurrence of such a biased argumentation.
    \item Since optimism bias and confirmation bias influenced the overall atmosphere of the meetings – we propose that, at the end of the meeting, after making the initial decisions, teams should explicitly list their drawbacks. Then, if the need arises, decisions should be changed accordingly.
    \item We will add an additional code to the coding scheme - ``decision''. Which will mean the decision that was ultimately made during the meeting. This will enable us to count how many rational and biased arguments were made in favour of the decisions that were eventually chosen.
    \item Instead of a simple debiasing presentation, we will hold a longer debiasing workshop. During this workshop, we will do more than simply inform the participants about the influence of cognitive biases on architectural decision-making. The participants will also be taught, through a series of practical exercises, how to apply our debiasing techniques.
    \item The next experiment will be performed on a significantly bigger sample of participants. 
\end{itemize}

\section{Conclusion}
The preliminary results (see Section \ref{results_section}) show that a simple presentation about cognitive biases and their possible influence on architectural decision-making is not an effective debiasing method. At the same time the pilot study revealed crucial information about how biases influenced the arguments for and against certain decisions. This made it possible to develop a series of modifications to our debiasing approach (as presented in Section  \ref{research_outlook_section}) in order to reshape the entire experiment.

\bibliographystyle{splncs04}
\bibliography{Debiasing-pilot.bib}










\end{document}